\documentclass[conference]{IEEEtran}
\IEEEoverridecommandlockouts

\usepackage{graphicx}
\usepackage{amsmath}
\usepackage[utf8]{inputenc}
\usepackage{color}
\usepackage{graphicx}
\usepackage{rotating}
\usepackage{tikz}
\usepackage{amssymb}
\usepackage{epsfig}
\usepackage{epstopdf}
\usepackage{algorithm}
\usepackage{algorithmic}
\usepackage{amsmath}
\usepackage{amsfonts}
\usepackage{dsfont}
\usepackage{url}
\usepackage{chemarr}
\usepackage{chemarrow}
\usepackage{bbold}
\usepackage[version=3]{mhchem}

\usepackage{mathabx}
\usepackage{tabularx,booktabs}
\usepackage{booktabs}
\usepackage{caption}
\usepackage[font=small,labelfont=bf]{caption}

\def\BibTeX{{\rm B\kern-.05em{\sc i\kern-.025em b}\kern-.08em
  T\kern-.1667em\lower.7ex\hbox{E}\kern-.125emX}}
  
 \usepackage[letterpaper, left=0.65in, right=0.65in, bottom=1in, top=0.75in]{geometry}

\begin{document}
\title{\huge{Analysis of Molecular Communications on the Growth Structure of Glioblastoma Multiforme }}

\author{Hamdan Awan, Andreani Odysseos, Niovi Nicolaou  and Sasitharan Balasubramaniam. %
% 
%\thanks{} 

}

\markboth{Submitted to IEEE Globecom 2021}{}
\maketitle

\begin{abstract}
In this paper we consider the influence of intercellular communication on the development and progression of \textit{Glioblastoma Multiforme} (GBM), a grade IV malignant glioma which is defined by an interplay Grow i.e. self renewal and Go i.e. invasiveness potential of multiple malignant glioma stem cells. Firstly, we performed wet lab experiments with U87 malignant glioma cells to study the node-stem growth pattern of GBM. Next we develop a model accounting for the structural influence of multiple transmitter and receiver glioma stem cells resulting in the node-stem growth structure of GBM tumour. By using information theory we study different properties associated with this communication model to show that the growth of GBM in a particular direction (node to stem) is related to an increase in mutual information. We further show that information flow between glioblastoma cells for different levels of invasiveness vary at different points between node and stem. These findings are expected to contribute significantly in the design of future therapeutic mechanisms for GBM.
\end{abstract}

%This paper is focused on understanding the mechanisms by which a network of multiple bio-nanomachines (transmitters and receivers) in a molecular communication system influence the growth structure of GBM tumour. By using a voxel model for multiple transmitter and receiver cells, this paper provides new insights into the role of inter-cellular communication (i.e. the mutual information between cells) in the evolution and progression of GBM (stem from node).  From the results of this paper we learn that the growth of tumour in any particular direction is driven by the increase in the mutual information between multiple cells in the node-stem structure. We further learn that information propagation speed between cells can vary at different points in the node and stem. This knowledge can be useful for developing future therapeutic mechanisms targeting GBM.

\IEEEpeerreviewmaketitle

\section{Introduction}
\label{sec:intro}
The inter-disciplinary research field of molecular communications bridges across different discipline such as  biological sciences and communication engineering formulating an  internet of nano-bio things \cite{akyildiz2015internet}. Molecular communication relies on information carrying signalling molecules as well as intercellular communications to encode the information to be transmitted. These molecules propagate through a medium by using diffusion \cite{kuscu2019transmitter} or active transport \cite{farsad2011simple}. The incoming molecules are captured at the receiver end, enabling the decoding of input signal \cite{Chou:2014jca}. The past decade has experienced a growing interest in applications of molecular communication in biomedical research. One such application is to study dynamic processes underlying the progression or growth of Glioblastoma Multiforme (GBM). GBM tumourigenesis relies on the dynamics of molecular communication networks in an interplay between ‘transmitter’ and ‘receiver’ glioma stem cells (GSCs) giving rise to GBM tumour invasion and progression. In this process the key role is performed by inter cellular communications between GSCs and Glioma Cells (GCs) \cite{jhaveri2016tumor}. 

In recent literature researchers have proposed a novel molecular communication based therapeutic mechanism to tackle GBM \cite{9083952}. In our previous work \cite{9398928} we proposed a simple voxel model to study the role of inter-cellular communication in the growth of GBM.  However, the influence of cell-cell interactions (communications) on the structural formation of GBM during their self-organisation process still remains an outstanding issue.  In order to develop an optimal therapeutic mechanism for GBM we aim to address this gap in knowledge.  In particular this paper aims to study the role of intercellular communication in the evolution of GBM structure as a molecular communication network. The specific contributions of this paper are: (a) Mathematical model for inter-cellular molecular communication for multiple transmitter and receiver GSCs (self-proliferating) and Glioma (invasive) cells resulting in GBM tumour growth. (b) To account for the self organization role of Glioblastoma cells to form a structure comprising of node and stem in a particular direction depending on the mutual information between malignant glioma cells which evolve to the highest grade tumour, i,e. GBM. By studying the impact of mutual information at various locations (i.e. between node and stem cell or within stem cells) we can utilise this knowledge to design future therapeutic mechanisms for GBM.

%   \begin{figure}
% \centering
% \includegraphics[trim=0cm 0cm 0cm 0cm ,clip=true, width=1\columnwidth]{zxc.png}
%  \caption{Glioblastoma Multiforme as a Molecular Communication Network: GSCs self-renewal and Glioma cells invasiveness function}
% \label{Gladiator2}
%  \end{figure}
 
The rest of the paper is organized as follows. In Section II we present the system model which includes the transmission and propagation mechanisms. Section III presents the complete molecular communication system model including diffusion and reactions. Next in Section IV we derive expressions for mutual information for multiple cells and follow it up with expressions of information propagation speed and GBM growth. Next in Sections V and VI we present numerical results and conclusion respectively.

 \begin{figure}
  \centering
 \includegraphics[trim=0cm 0cm 0cm 0cm ,clip=true, width=0.8\columnwidth]{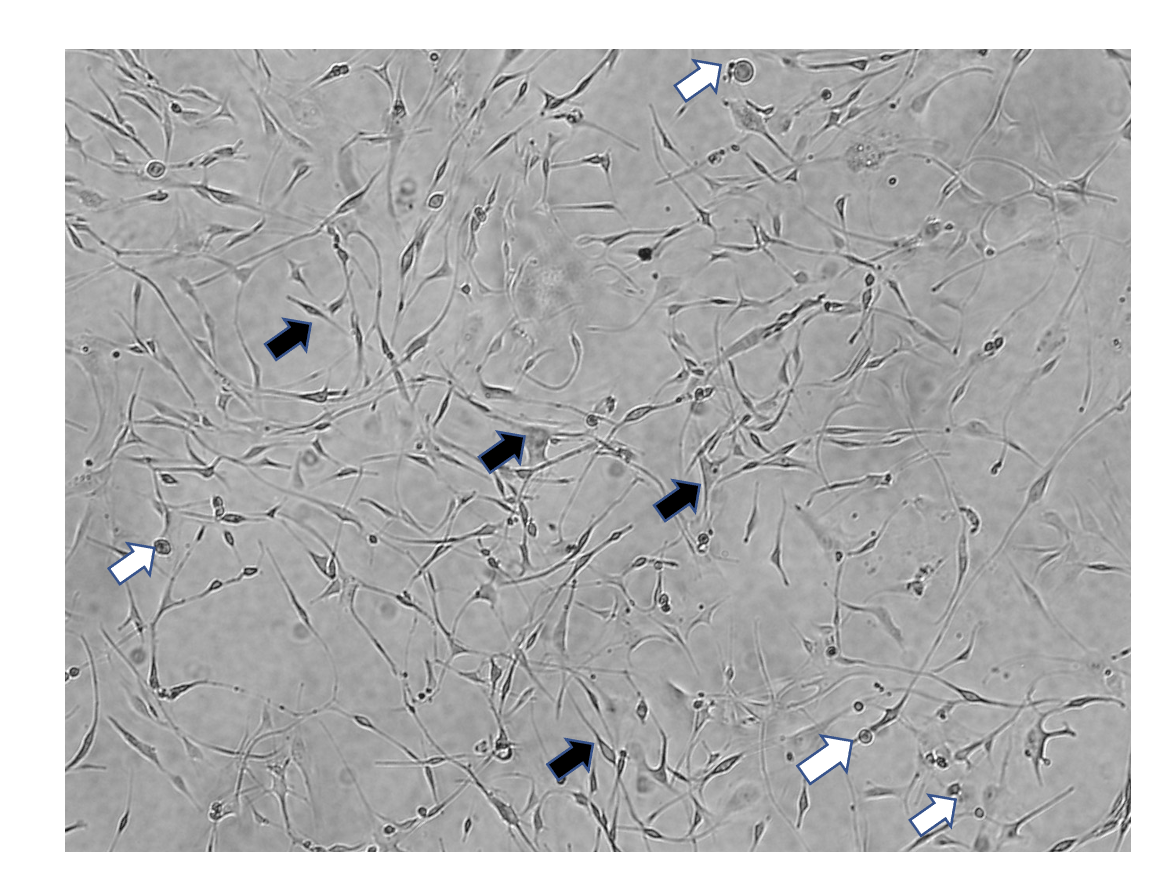}
  \caption{Grade IV U87 Malignant Glioma Cells in culture- White arrows represent Grow (GSC)  and black arrows represent Grow (malignant GC).}
\label{system parallel voxel}
\end{figure}
 
 \begin{figure}
  \centering
 \includegraphics[trim=0cm 0cm 0cm 0cm ,clip=true, width=1.1\columnwidth]{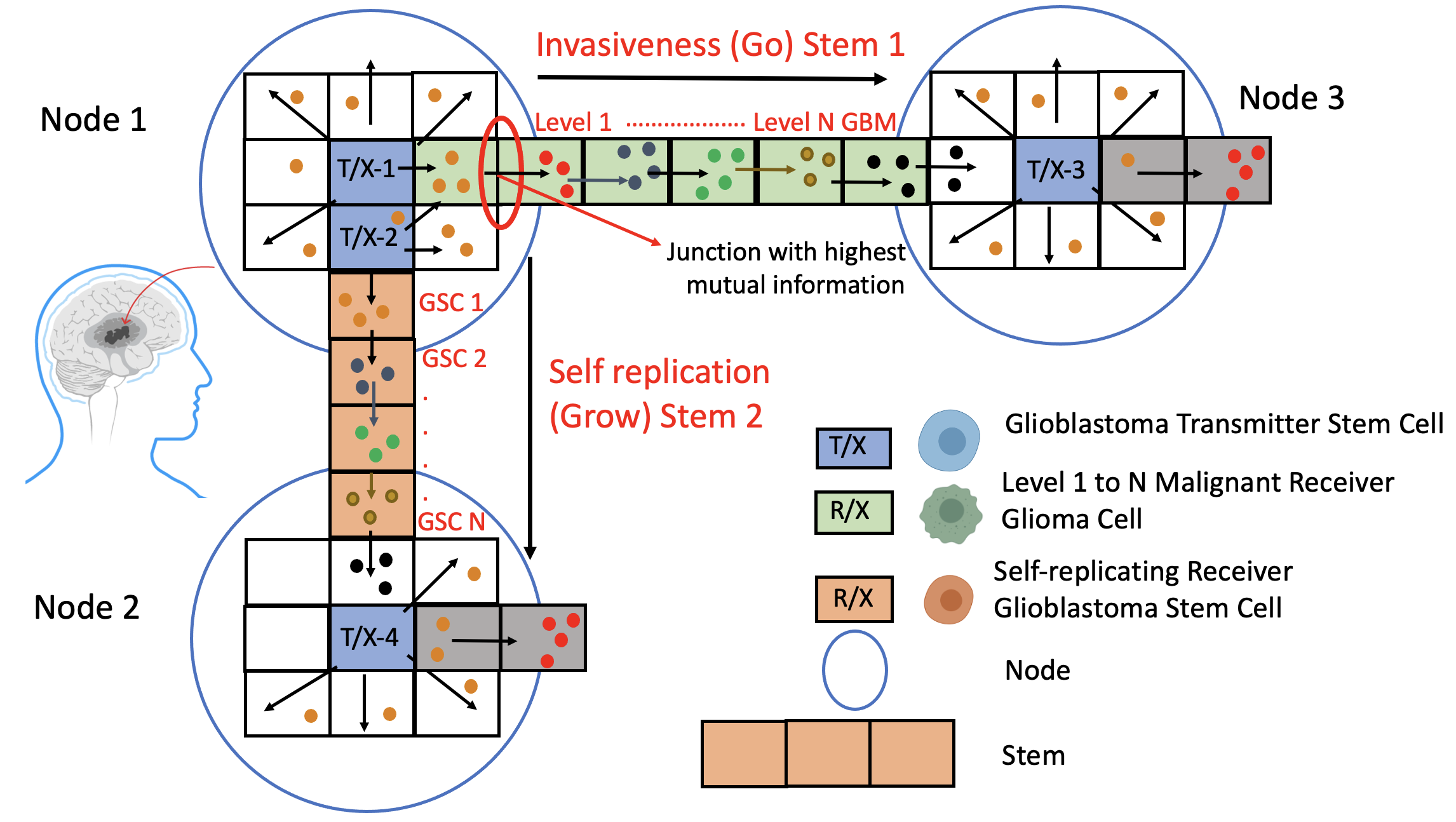}
  \caption{Evolution of Go and Grow based node-stem structure of GSC cells leading to GBM growth.}
\label{system parallel voxel 2}
\end{figure}

 \section{System Model}
In this section we present the system model accounting for the influence of multiple transmitter and receiver glioma cells on the structure of GBM. From literature we learn that errors in internal signalling of neurons result in the generation of a glioma stem cell (GSC) \cite{jung2019emerging}. This GSC acts as an initiator to generate and progress GBM tumour by using inter-cellular molecular communication. Figure 1 shows the wet lab experimental results for the growth pattern of U87 malignant glioma cells. The white arrows in Figure 1 show spheroid cells representing Glioma Stem Cells whereas black arrows show malignant glioma cells (GBM) at different levels of invasiveness. For both these cases we observe a node-stem structure associated with growth of GBM. In this paper we present a theoretical model to explain these experimental results where a higher mutual information between multiple cells result in this node-stem growth structure of GBM. Figure 2 shows GBM growth as a cell signalling network comprising of multiple transmitter and receiver malignant glioma cells which leads to GBM tumour growth through: (a) self-renewing GSCs i.e. Grow and (b) invasiveness from  level $1$ to level $N$ GBM i.e. Go (c) Hybrid of Grow and Go.  Figure 1 shows the self-organization of malignant cells to form a node-stem structure (using inter-cellular communication). The transmitter glioma stem cells T/X-1 and T/X-2 (i.e. node) emit signalling molecules that diffuse through propagation medium to the neighbouring cells in different directions. However, the self-replication (i.e. Grow configuration) up to $N$ GSC receiver cells in one direction (i.e. stem growth) depends on the mutual information at each junction (i.e. boundary with neighbouring cell/voxel) leading to the increased growth of GBM in the direction of higher information flow. Similarly we can consider the invasiveness of GSC cells (i.e. Go configuration)  from level $1$ to level $N$ GBM increasing more in one particular direction depending on the mutual information at the junction.

%Overall, we consider three separate configurations of cells leading to GBM growth:  (a) series, corresponding to self renewal based stem structure or \textit{Grow} potential as shown in Figure 1 (b) series, corresponding to invasiveness based stem structure or \textit{Go} potential similar to Figure 1, and (c) hybrid of both self-renewal and invasiveness leading to the structural formation of GBM. The information driving the \textit{Grow} potential, for example, is transferred from GSC-1 i.e. node transmitter cell T/X-1 emitting information resulting in proliferation and maintenance of the stem structure potential to GSC-1 receiver cell through diffusing molecules. In the next step the GSC-1 acts as the transmitter for the GSC-2 receiver cell and the stem growth process continues up to $N$ number of cells resulting in development of the GBM tumour

% \begin{figure*}
% \centering
% \includegraphics[trim=0cm 0cm 0cm 0cm ,clip=true, width=0.70\textwidth]{zyc.png}
%  \caption{Initiation and progression of Glioblastoma Multiforme - An interplay between Go and Grow phenotypes. Go, denotes evolution of progressively invasive phenotypes. Grow, denotes maintenance of self-renewal and stemness phenotypes. Configuration or phenotype series: (A) Invasiveness or Go; (B) Stemness/self-renewal or Grow; (C) Hybrid cell propagation.}
% \label{Glad}
%  \end{figure*}
 
Figure 1 shows the node-stem structure for both Grow and Go configurations for multiple transmitter cells. The input information molecules at each step (cell) along the length of stem propagate to the receiver malignant glioma cells through diffusion leading to chemical reactions which produce output molecules.  In this work we study how varying number of transmitter and receiver cells can influence the node-stem structure of GBM. By using mathematical modelling we show that the GBM growth rate in different directions depend directly on the mutual information between malignant glioma cells. Note that the example presented is for two transmitter cells, however it is straightforward to extend this general model to incorporate increasing number of transmitters cells.

\subsection{Transmitter and Propagation Mechanism}
In order to model the transmission and propagation we use a previously developed voxel based model for intercellular molecular communication  \cite{awan2019molecular,riaz2020using}. Each transmitter is assumed to occupy a single voxel emitting input molecules $P$ at a rate of $U_i(t)$. For multiple transmitter cells the different levels of input signals correspond to different input concentrations of molecules.  Considering example in Figure 1, the information carrying molecules emitted by the transmitter diffuse through a propagation medium to the neighbouring receiver cells resulting in tumour growth due to Go diffusion and Grow diffusion. Likewise, the propagation medium is assumed to be divided into a number of  voxels \cite{awan2016demodulation}. Figure 2 shows the 3-d hybrid configuration of the voxel model depicting intercellular communication leading to the node-stem structure of the GBM tumour corresponding to different levels of invasiveness or self-renewal of malignant glioma cells.  

%  Figures 3 and 4 show examples of a 2-dimensional (2-d) propagation medium consisting of 5$\times$1$\times$1 voxels for \textit{Go} and  4$\times$1$\times$1 voxels for \textit{Grow} configurations, respectively. Figure 2 represents the case where we consider GBM tumour growth solely attributed to the contribution of the malignant glioma cell invasive potential corresponding to different stages of invasiveness up to formation of GBM. Similarly, Figure 3  represents the case where we consider GBM tumour growth solely attributed to the contribution of self-renewal or proliferation of the GSCs. The different colours of molecules correspond to reactions at each receiver cell producing output molecules which act as the input to the next cell in the chain. Notably, for simplicity and easier perception of the model each voxel is indexed to represent distinct cells such as GSC-1 to GSC-4. 

To explain the propagation of input molecules we use the concept of spatially discrete jumps accounting for inter-voxel movement of molecules. This is modelled for both Grow  (self-replication) diffusion and Go (invasive) diffusion configurations such that movement of molecules in one particular direction leads to the development of stem like structure of GBM as shown in Figure 1. For this paper we assume a homogeneous medium with a constant diffusion coefficient $D$. By applying a finite difference discretization to the 3-d diffusion \cite{gardiner2009stochastic} we can obtain the probability of the jumping of single molecule between two voxels i.e $\frac{D}{d^2}$ where $d$ is voxel edge length. Furthermore, this model can account for both reflection and absorption boundary conditions.

  \begin{figure}
  \centering
\includegraphics[trim=0cm 0cm 0cm 0cm ,clip=true, width=0.75\columnwidth]{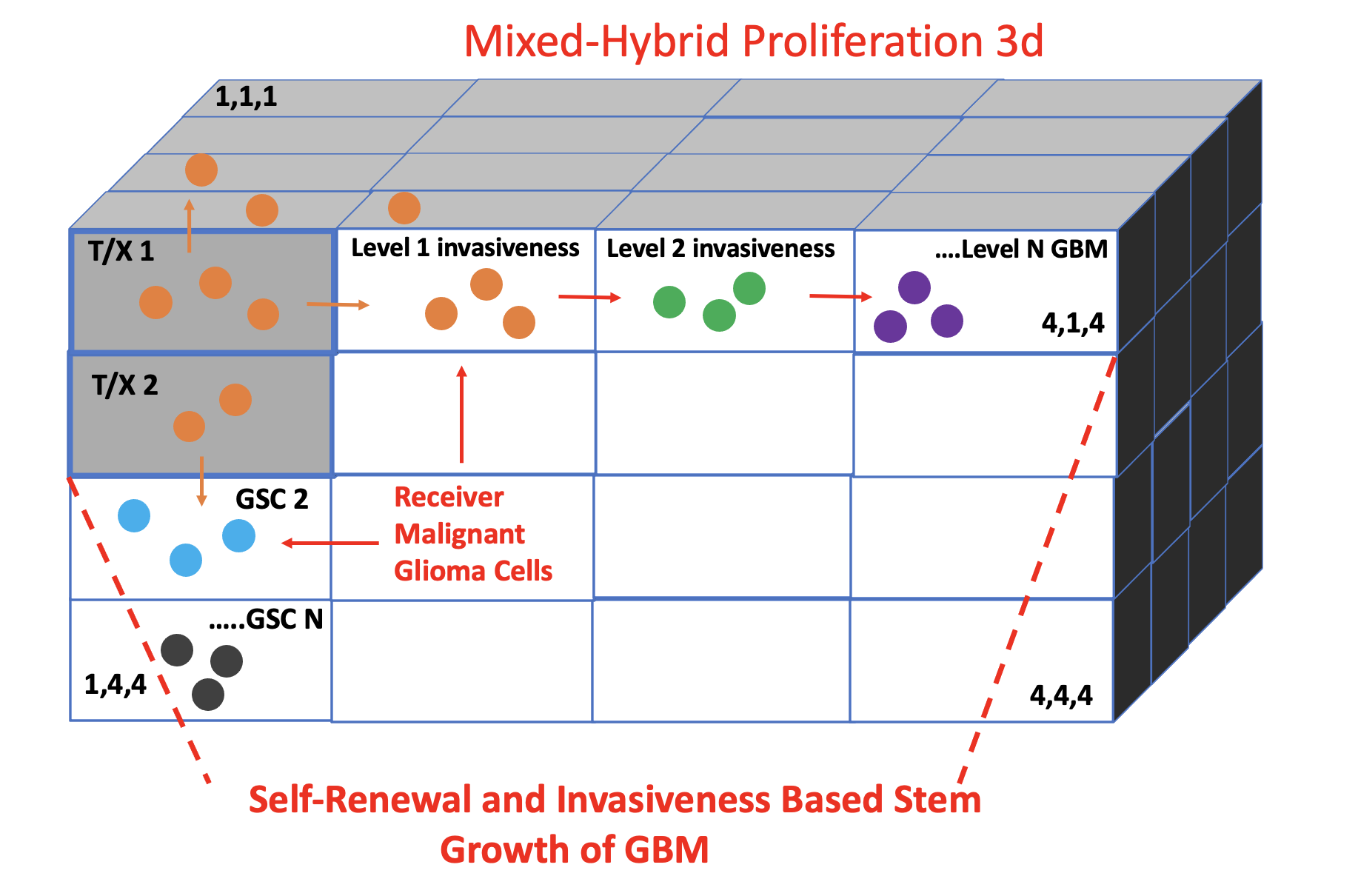}
 \caption{3-dimensional voxel model Hybrid Configuration}
\label{system mixed voxel}
 \end{figure}

\begin{figure}
\centering
\includegraphics[trim=0cm 0cm 0cm 0cm ,clip=true, width=0.75\columnwidth]{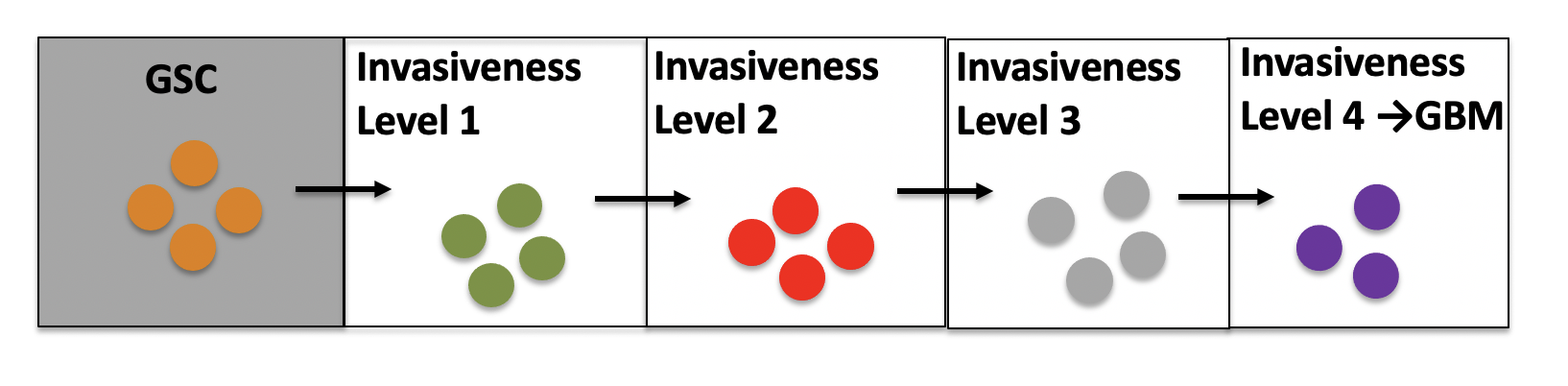}
 \caption{Series Voxel Model of \textit{Go} Configuration}
\label{system series voxel}
 \end{figure}

\subsection{Diffusion Reaction System}
\subsubsection{\textbf{Diffusion Subsystem}}

For this paper assume a general 3-d propagation medium of size $M_x \times M_y \times M_z$ cubic voxels. Each cubic voxel is assumed to represent a single transmitter or receiver cell. Figure 2 shows a 3-d dimensional hybrid configuration of malignant glioma, or glioblastoma, cells with $M_x = M_y = M_z = 4$. For the ease of understanding we explain a simple example of voxel model in Figure 3 which depicts the Go (invasiveness) configuration of cells i.e. $M_x = 5$ and $M_y = 1 $ and $M_z$ = 1.   For this model, we obtain $n_{P,i}$ i.e. the number of input molecules $P$ in voxel $i$ as:
\begin{equation}
n_P (t) = [n_{P,1}(t),n_{P,2}(t),n_{P,3}(t) ,n_{P,4}(t)]^T 
\label{1q:a}
\end{equation}
The terms on the R.H.S of Eq. \eqref{1q:a} are updated with each new diffusion event. For example for the diffusion of a single molecule from transmitter cell-1 to receiver malignant glioma cell, the term $n_{P,1}(t)$ decreases by 1 whereas the term $n_{P,2}(t)$ increases by $1$. This is explained by using a jump vector $q_{d,1} (t) = [-1 ,1, 0, 0]^T$. The new state of the system after this diffusion event becomes $n_P (t)+q_{d,1}$. Additionally the term $W_{d,1}(n_P (t))= dn_{P,1}$ represents the jump rate function for this diffusion event. By combining all these jump vectors and the corresponding jump rate functions corresponding to each diffusion event we obtain a diffusion matrix $H$ as in \cite{awan2019molecular}:
\begin{align}
H =  
\left[ \begin{array}{ccccc}
-d & 0 & 0 & 0  \\
d & -d & 0 & 0  \\
0 & d & -d & 0   \\
0 & 0 & 0 & d  \\
\end{array} \right] 
\label{eqn:H} 
\end{align}    
where $-d$ indicates diffusion out of a voxel. The size of matrix $H$ depends on the number of interacting cells in the system. Next we use a SDE to model all the diffusion events \cite{gardiner2009stochastic}: 
\begin{align}
& \dot{n}_P(t) = \sum_{j = 1}^{J_d} q_{d,j}W_{d,j} (n_P(t))   + \sum_{j = 1}^{J_d} q_{d,j} \sqrt{W_{d,j}( n_P (t))} \gamma_j \nonumber \\ & + {\mathds 1}_T U(t).
\label{eqn:sde:do} 
\end{align}
Eq. \eqref{eqn:sde:do} is therefore a form of a chemical Langevin equation where the term $\sum_{j = 1}^{J_d} q_{d,j}W_{d,j} (n_P(t))$ accounts for the deterministic dynamics of the system and $\gamma_j$ represents continuous-time Gaussian white noise. If all the jump rates are assumed as linear this leads to:
\begin{align}
H n_P(t) & = \sum_{j = 1}^{J_d} q_{d,j}W_{d,j} (n_P(t)).  
\end{align}

The term $ \sum_{j = 1}^{J_d} q_{d,j}W_{d,j} \sqrt {W_{d,j}(n_P(t))}$ accounts for the  the stochastic dynamics of the system. The term ${\mathds 1}_T U(t)$ represents the transmitter input, i.e. emission rate.

 \subsubsection{\textbf{Reaction Subsystem}}
\label{reaction}
To derive the SDE for the reaction sub-system we assume that a ligand-binding type reaction takes place at each step of the node-stem structure of cells. In first reaction (5) the input  molecules  $P$ released from transmitter malignant glioma cell T/X-1 propagate to level 1 invasive receiver cell to produce output molecules $X_{[1]} $. These $X_{[1]}$ molecules act as the input in the next reaction (6) resulting in production of level 2 invasiveness $X_{[2]}$ molecules. This process  process continues up to cell $N$ resulting in Level $N$ GBM.
\begin{align}
\cee{
P &<=>C[k_1][k_0] X_{[1]}, \;  \;\;   k_1 n_{P,R},  \;\; k_0 n_{X_1} \label{cr:rc1} \\
X_{[1]}  &->C[k_2] X_{[2]}, \; \;\; k_2 n_{X_1} \label{cr:rc2} \\
X_{[N-1]} &->C[k_N] X_{[N]}, \; \;\; k_N n_{X_N}  \ 
 \label{eq:numd}  }
\end{align}

Where $n_{P,R}$ represents the number of input molecules in the receiver glioma cell  $i$ and $n_{X_i}$ represent the number of output molecules in receiver glioma cell $i$. The symbols $k_0$ - $k_N$ are the respective reaction rate constants. In Eq. \eqref{cr:rc1} the input molecules $P$ react at the rate ${k}_{1} n_{P,R}$ producing output molecules $X_{[1]}$. 
For the sake of simplicity we derive the SDE for the case when we have only one receiver reaction i.e.Eq. \eqref{cr:rc1}. The input and output for this case is $ n_{P,R} $ and $ n_{X_1} $ respectively. We obtain the state vector as:

\begin{align}
 \tilde{n}_R(t) 
 & =  \left[ \begin{array}{c|c}
 n_{P,R}(t) & n_{X_1}(t)  
\end{array} \right]^T
\end{align} 

and the corresponding SDE is given as:

\begin{align}
\dot{\tilde{n}}_R(t) & = {\rm I\!R} \tilde{n}_R(t) + \sum_{j = J_d+1}^{J_d + J_r} q_{r,j} \sqrt{W_{r,j}(\langle \tilde{n}_R(t) \rangle)} \gamma_j. 
\label{eqn:sde:ro11} 
\end{align}

where $q_{r,j}$ and $W_{r,j}$ represent jump rates and jump vectors for the reaction 
events.  The term $J_d$+ $J_r$ represents the total number of events (diffusion and reactions). \textcolor{black} {
${\rm I\!R}$ is 2$\times$2 matrix  with entries depending on reaction  \eqref{cr:rc1} see Table 1.}

% Note that this derivation of SDE is for a single glioma receiver cell with reactions described in \eqref{cr:rc1}. However, when we consider a higher number of malignant glioma receiver cells in the system this will lead to the derivation of a number of SDEs accounting for the output of each malignant glioma receiver cell. Note that these SDEs will essentially have the same form as Eq. \eqref{eqn:sde:ro11} with an updated $R$ matrix. 

\begin{table}
\centering
\caption{${\rm I\!R}$ Matrix for receiver Reactions}
\begin{tabular}{|c|c|}
\hline
\multicolumn{1}{|c|}{Receiver }	&	\multicolumn{1}{|c|}{${\rm I\!R}$ Matrix}	\\
\hline
Reversible Reaction 5 &    $ \begin{bmatrix}  -k_{1} & k_{0} \\ k_{1} & -k_{0} \end{bmatrix}$
\\ \hline
\end{tabular}
\label{table:1a}
\end{table}

\section{End to End Model}
\label{complete}
 To obtain the stochastic differential equation for the end to end model we combine the SDEs for diffusion-only and reaction-only subsystems. In this paper we have separately derived SDE's for each subsystem in order to study the interconnection of both the subsystems. As the term  $n_{P,R}(t)$ is common in the expressions for both $n_P(t)$ and $\tilde{n}_R(t)$ as shown in Eq.s (3) and (9), we consider it as the  interconnection point between the two subsystems. For diffusion subsystem in Figure 3, the dynamics of number of signaling molecules in the receiver malignant glioma cell is given as: 
 \begin{align}
& \dot{n}_{P,R}(t) = d n_{P,T}(t) - d n_{P,R}(t) \nonumber \\
 & + \underbrace{\sum_{j = 1}^{J_d} [q_{d,j}]_R \sqrt{W_{d,j}(n_P(t))} \gamma_j}_{\xi_d(t)}
 \label{eqn:sde:ro:r2}  
 \end{align}
 where $n_{P,T}$ represents the number of input molecules diffusing in the receiver. $[q_{d,j}]_R$ represents $R$-th element of jump vector $q_{d,j}$. Similarly for receiver subsystem, the dynamics of the input signalling molecules as a result of reaction  (5) is given by the first element of Eq.~\eqref{eqn:sde:ro11}, i.e.: 
 \begin{align}
& \dot{n}_{P,R}(t) = R_{11} n_{P,R}(t) + R_{12} n_{X_1}(t)  \nonumber \\
&+ \underbrace{\sum_{j = J_d+1}^{J_d+J_r} [q_{r,j}]_1 \sqrt{W_{r,j}(\tilde{n}_R(t))} \gamma_j}_{\xi_r(t)} 
\label{eqn:sde:ro:r1a} 
\end{align}
 where $[q_{r,j}]_1$ is the first element of jump vector $q_{r,j}$.
We can then obtain the SDE for the end to end system by combining separate SDE's derived in Eqs. \eqref{eqn:sde:ro:r2} and \eqref{eqn:sde:ro:r1a}: 
% \begin{align}
% \dot{n}_{P,R}(t) = &  d n_{P,T}(t) - d n_{P,R}(t) + R_{11} n_{P,R}(t)  + R_{12} n_{X_1}(t) \nonumber \\
% & + \xi_{total}(t) 
% \label{eqn:sde:nlr:ex} 
% \end{align}
% where $\xi_{total}(t) = \xi_d(t) + \xi_r(t)$ accounts for the total noise in the system, i.e., sum of the noise due to the diffusion and reaction events. The state of the end to end system is obtained by: 
% \begin{align}
% n(t) = 
%  & \left[ \begin{array}{c|c}
%  n_{P}(t)^T & n_X(t)  
% \end{array} \right]^T.
% \label{eqn:state} 
% \end{align}
This shows that we can obtain SDE for the end to end system is obtained by combining SDE's derived in Eq.s (3) and (9)) as:
 \begin{align}
 \dot{n}(t) & = A n(t) + \sum_{j = 1}^{J} q_j \sqrt{W_j(\langle n(t) \rangle)} \gamma_j + {\mathds 1}_T U(t). 
 \label{eqn:mas11a}
 \end{align} 
where $\langle n(t) \rangle$ (i.e. the mean of $n(t)$) is the solution to the following ODE:
 \begin{align}
 \dot{\langle n(t) \rangle} & = A \langle n(t) \rangle+ {\mathds 1}_T c 
\label{eqn:ode_ninfinity}
\end{align} 

 \textcolor{black}{We can obtain the number of input molecules in the receiver (by taking Laplace of Eq. \eqref{eqn:mas11a}) as:}
 \begin{align}
  N(s)  = 
  \underbrace{   (sI - A)^{-1} {\mathds 1}_T }_{\Psi(s)}   U(s).
 \label{eqn:mas11}
 \end{align}

Multiplying this expression with $ 1_{X_1}$ we obtain the number of output molecules in receiver cell-1:
\begin{align}
  N_{X_1}(s)  = 
 1_{X_1}  \underbrace{  (sI - A)^{-1} {\mathds 1}_T }_{\Psi(s)}   U(s)
\label{eqn:mas11b}
\end{align}
% \begin{align}\mbox{where,  }
% \Psi(s) = \frac{G_{{X_1}L} (s) H_{RT} (s)}{1 - (R_{11} + G_{LL} (s) )H_{RR} (s)} 
% \end{align}

% in which,

% \begin{align}
% & H_{RT}(s) = {\mathds 1}_R^T (sI - H)^{-1} {\mathds 1}_T , \nonumber \\
% & G_{PP}(s) = R_{12}(sI - R_{22})^{-1} R_{21} \nonumber \\
% & G_{{X_1}P}(s) = {\mathds 1}_X^T(sI - R_{22})^{-1} R_{21} ,\nonumber \\
% & H_{RR}(s) = {\mathds 1}_R^T (sI - H)^{-1} {\mathds 1}_R
% \end{align}
% The expressions $H_{RT}(s)$ is the Laplace transform of the probability that input molecule $P$ released by the GSC-1 at time $t=0$ arrives in first receiver malignant glioma cell at time $t$. $H_{RR}(s)$ is the Laplace transform of the probability that input molecule $P$ is present in the receiver malignant glioma cell 1 at time $t= 0 $ is still present in receiver malignant glioma cell 1  at time $t$.  $G_{{X_1}P}(s)$ and $G_{PP}(s)$ correspond to reactions in the system. The $G_{{X_1}P}(s)$ is the Laplace transform of the probability that an output molecule $X_{1}$ at time $t$ is produced by the input molecule $P$ at time $t=0$, whereas $G_{PP}(s)$ is the Laplace transform of the probability that input molecule $P$ in the receiver at time $t$ is produced by a input molecule $P$ in the receiver at time $t=0$ through the reactions in the receiver. This means that the transfer function $\Psi(s)$ effectively takes into account all the possible diffusion and reaction events in the system.

\section{Communication Properties of System}
\subsection{Mutual Information}
 \textcolor{black}{To derive expression for mutual information between the general input rate of molecules $U_i(t)$  and the general number of output molecules $n_{X_i}(t)$ i.e. $I(n_{X_i},U_i)$ (where $i=1,2,..n$ number of transmitters or receivers) we use an expression for two Gaussian distribution random processes from \cite{tostevin2010mutual}: }
% %
 \begin{align}
I(n_{X_i},U_i) &= \frac{-1}{4\pi} \int_{-\infty}^{\infty} \log \left( 1 - \frac{|\Phi_{n_{X_i}U_i}(\omega)|^2}{\Phi_{n_{X_i}n_{X_i}}(\omega) \Phi_{U_iU_i}(\omega)}  \right) d\omega
\label{eqn:MI0}
\end{align} 
% %
where $\Phi_{n_{X_i}n_{X_i}}(\omega)$ (resp. $\Phi_{U_iU_i}(\omega)$) represent the power spectral density of $n_{X_i}(t)$ ($U(t)$), and $\Phi_{n_{X_i}U_i}(\omega)$ is the cross spectral density of $n_{X_i}(t)$ and $U_i(t)$. From \cite{warren2006exact} we learn that that if all the jump rates $W_j(n(t))$  are linear in $n(t)$, then the power spectral density of $n(t)$ can be obtained by using Eq. \eqref{eqn:mas11}. We can therefore describe the dynamics of the system in Eq. \eqref{eqn:mas11a} by a set of linear SDE's with $U_i(t)$ as the general input  and $n_{X_i}(t)$  as the output. $\gamma_j$ in Eq.~\eqref{eqn:mas11a} accounts for the noise in the output. Therefore, Eq.~\eqref{eqn:mas11a} models a continuous-time linear time-invariant (LTI) stochastic system subject to a Gaussian input and Gaussian noise. We can then obtain the power spectral density $\Phi_{n_{X_i}}(\omega)$ as: 
 \begin{align}
 \Phi_{n_{X_i}}(\omega) & =  |\Psi(\omega) |^2 \Phi_{U_i}(\omega) + \Phi_{\eta}(\omega)
\label{2331a} 
 \end{align}
 where $\Phi_{U_i}(\omega)$ is the power spectral density of input $U_i(t)$ and $|\Psi(\omega)|^2$ is channel gain with $\Psi(\omega) = \Psi(s)|_{s = i\omega}$ defined as:
 \begin{align}
 \textcolor{black}{\Psi(s) = {\mathds 1}_{X_i}   (sI - A)^{-1} {\mathds 1}_T}   
 \label{21a}
 \end{align}
which is obtained by mean and Laplace of Eq.~\eqref{eqn:mas11a}. $\Phi_{\eta}(\omega)$ represents the stationary noise spectrum: 
\begin{align}
\Phi_{\eta}(\omega) & =   \sum_{j = 1}^{J_d + J_r} | {\mathds 1}_{X_1} (i \omega I - A)^{-1} q_j |^2 W_j(\langle n_{}(\infty) \rangle) 
\label{eqn:spec:noise2} 
\end{align} 
where $\langle n_{}(\infty) \rangle$ is the mean state of the system at time $\infty$ due to a constant input $c$. By using standard results for LTI system we obtain the cross spectral density $\Phi_{n_{X_i}U_i}(\omega)$ as:
\begin{align}
|\Phi_{n_{X_1}U}(\omega)|^2 &= |\Psi(\omega) |^2 \Phi_U(\omega)^2. 
\label{eqn:csd} 
\end{align} 

The final expression for mutual information $I(n_{{X_i}},U_i)$ is obtained by substituting Eqs.~\eqref{2331a} and ~\eqref{eqn:csd} in Eq.~\eqref{eqn:MI0} as:
\begin{align}
I(n_{{X_i}},U_i) = \frac{1}{2} \int \log \left( 1+\frac{ | \Psi(\omega) |^2}{\Phi_{\eta}(\omega)} \Phi_{U_i}(\omega) \right) d\omega .
\label{eqn:mi1ss}
\end{align}
Maximum mutual information can be obtained by applying the water-filling solution to Eq. \eqref{eqn:mi1ss}.

\subsection{Information Propagation Speed}
To compute the information propagation speed for multiple number of transmitter and receiver malignant glioma (or GBM) cells in different configurations we use the results of mutual information. For this we use the results of mutual information of the system for varying number of transmitters and receivers by using the expressions similar to Eq. \eqref{eqn:mi1ss}. Next, by selecting a suitable threshold value of mutual information, we calculate the time difference when mutual information curve for various cases cross this threshold value. Finally, we calculate the information propagation speed $V$ by using the following expression:
\begin{equation}
  V=  \frac{1}{\mathbf{E} [\Delta t_{i,i+1}]}
  \label{infop}
\end{equation}
where $\mathbf{E}$ is expectation operator and $\Delta t_{i,i+1}$ represents the time difference at which the mutual information for each case crosses the threshold.

\subsection{Mathematical Model for tumour Growth}
The GBM tumour growth relies on the cell proliferation probability $p(D)$ and accounts for both self-replicating  and invading invasive malignant glioma cells. To compute the growth of tumour we use the expressions derived in \cite{9398928} where the key variables controlling GBM growth are: the initial tumour growth rate $\lambda(0)$, the tumour growth rate over time $\lambda(t)$, the vascular growth retardation factor $\theta$ and the volume of growing tumour $V_T(t)$ (proportional to the number of receiver cells producing output molecules). \textcolor{black}{The tumour volume increases  according to the following ODE's:}

\begin{equation}
    \frac{dV_T}{dt} = \lambda(t)V_T \approx n_{{X_i}}(t)V_T
    \label{eq:tum}
\end{equation}

\begin{equation}
    \frac{d\lambda}{dt} = -\theta \lambda(0)\lambda.
     \label{eq:tum2}
\end{equation}

% % We consider the volume of tumour $V_T$ as proportional to the number of tumour cells with each tumour cell dividing in a specific cell cycle time, $T_{cc}$.

% % The function $f$ is the growth rate of the dividing tumour cell and depends on both the total volume of actively dividing cells, $V_T$, and the tumour vascular volume, $V_v$. The function $g$ is the transition rate of the dividing cells to the non-dividing cells.

% % \begin{Equation}
% %     \frac{d V_T}{dt} = f(V_T,V_v)V_T - gV_T
% % \end{Equation}
 
% % where 
% % \begin{Equation}
% %      f(V_T,V_v) = \lambda(V_T,V_v) p(D)
% % \end{Equation}
 
where $V_T$ represents the volume of tumour depending on  the cell proliferation probability. With the variations in the inter-cellular communication leading to the increasing flow of information molecules the GBM tumour volume increases. In particular, the tumour growth rate  $\lambda(t)$ depends directly on the output of molecular communication system $n_{{X_i}}(t)$. $\theta$ $\leq$ 1 represents the retardation factor of the vascular structure.

%  \begin{figure}
% \begin{center}
% \includegraphics[trim=0cm 0cm 0cm 0cm ,clip=true, width=0.9\columnwidth]{f7.eps}
% \caption{ Mutual information vs increasing number of Receiver cells}
% \label{molecular communications-n3}
% \end{center}
% \end{figure}

% \begin{figure}
% \begin{center}
% \includegraphics[trim=0cm 0cm 0cm 0cm ,clip=true, width=0.9\columnwidth]{f8.eps}
% \caption{Mutual information per cell vs increasing number of Receiver cells}
% \label{molecular communications-n4}
% \end{center}
% \end{figure}

% \begin{figure}
% \begin{center}
% \includegraphics[trim=0cm 0cm 0cm 0cm ,clip=true, width=0.9\columnwidth]{f9.eps}
% \caption{Information Propagation Speed for Receiver cells in series-\textit{Go} Configuration}
% \label{Result 9}
% \end{center}
% \end{figure}

% \begin{figure}
% \begin{center}
% \includegraphics[trim=0cm 0cm 0cm 0cm ,clip=true, width=0.9\columnwidth]{f10.eps}
% \caption{Information Propagation Speed for Receiver cells in series-\textit{Grow} Configuration}
% \label{Result 3}
% \end{center}
% \end{figure}

\section{Numerical Results and Discussions}
In this section we present numerical results for different configurations and multiple malignant glioma transmitter and receiver cells on the structure of GBM growth. Some of the main parameters considered in this work are based on previous work \cite{9398928}. The propagation medium is assumed as rectangular prism which is divided into a number of cubic voxels. For example if we assume a a propagation medium of size 1$\frac{2}{3}\mu$m$\times$1$\frac{2}{3}\mu$m$\times$1$\frac{2}{3}\mu$m where each voxel has a size $\frac{1}{3}$ $\mu$m$^{3}$ this will lead to rectangular prism of $5 \times 5 \times 5$ voxels. Using this general methodology we can vary the propagation medium sizes for different system settings such as the number of cells or input molecules. For neural cells considered in this paper we consider size = 20 $\mu$m  \cite{diao2019behaviors}.  We further assume diffusion coefficient $D$ = 1 $\mu$m$^2$s$^{-1}$ for the propagation medium. The reaction rate constants are  $k_1$ = $k_0$ =  1. For each cell we assume six neighbouring cells as in 3-d voxel model with the distance  between cells as 1-5 $\mu$ m. The radius of the propagation environment is 4  mm and cell proliferation probability $p(D)$ is assumed to be  1. The tumour growth retardation factor $\theta$  is 0.53-0.99 whereas the tumour doubling time is  $0.693/\lambda$.  For the results presented in this section we assume absorbing boundary condition such that molecules can escape the medium at rate $\frac{d}{50}$.  The value of deterministic emission rate $c$ is chosen such that it derives the transformation of transmitter GSC cell into a tumour cell. Our main aim is to explore the impact of multiple transmitter cells on the mutual information and hence the structure of GBM tumour growth for varying number of transmitter and receiver cells in different configurations. The main results obtained in this paper are as follows. First in Figure 5 we present the results for growth rate of tumour which proliferates by receiving input information carrying molecules emitted by the transmitter. We show that with the increase in the number of input  molecules the growth of tumour at time $t$ also increases. However we observe that the growth is higher in hybrid configuration.

 \begin{figure} [ht]
 \begin{center}
\includegraphics[trim=0cm 0cm 0cm 0cm ,clip=true, width=0.8\columnwidth]{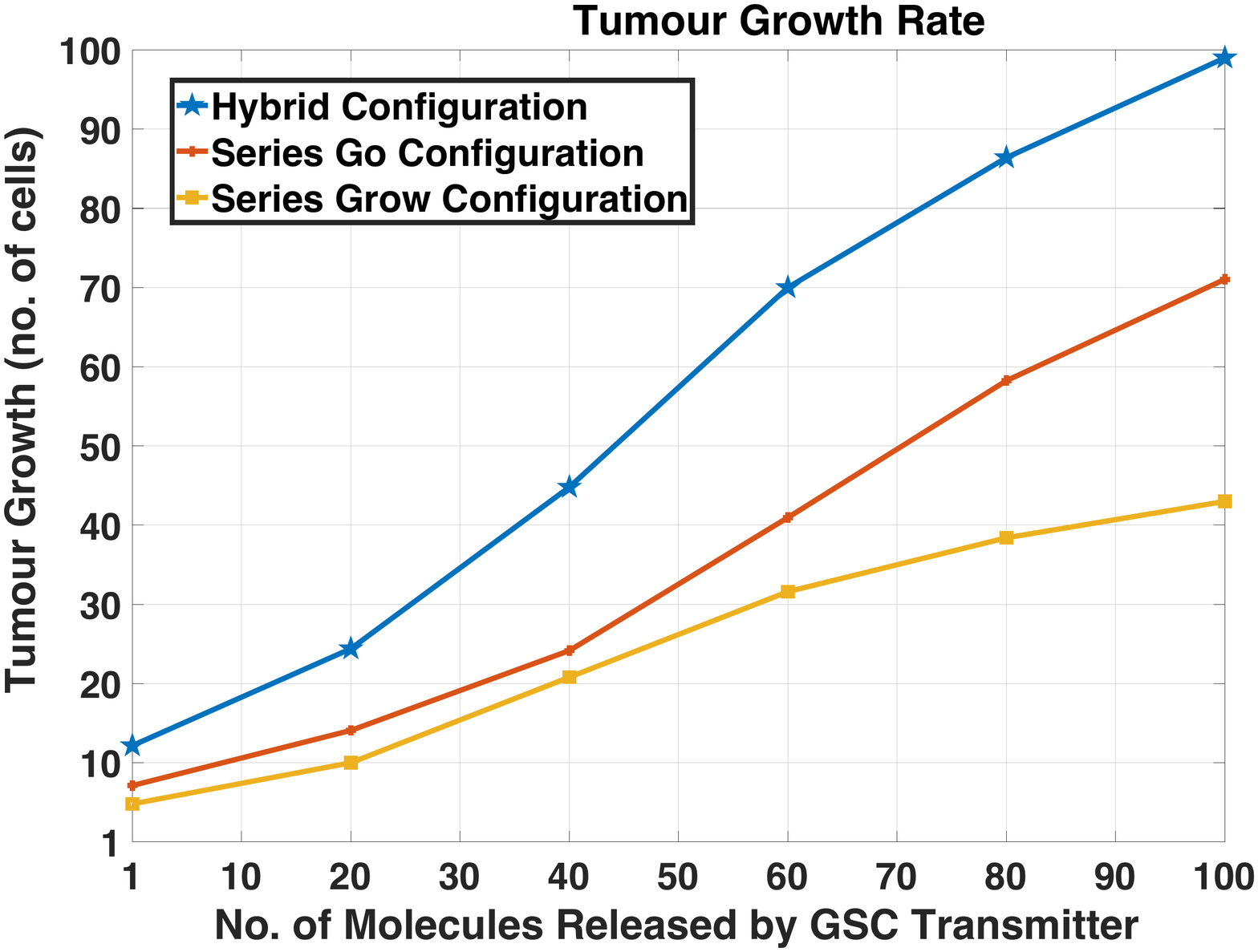}
 \caption{Tumour Growth rate (number of cells) vs No. of molecules }
\label{s2}
 \end{center}
 \end{figure}

  \begin{figure} [ht]
 \begin{center}
\includegraphics[trim=0cm 0cm 0cm 0cm ,clip=true, width=1\columnwidth]{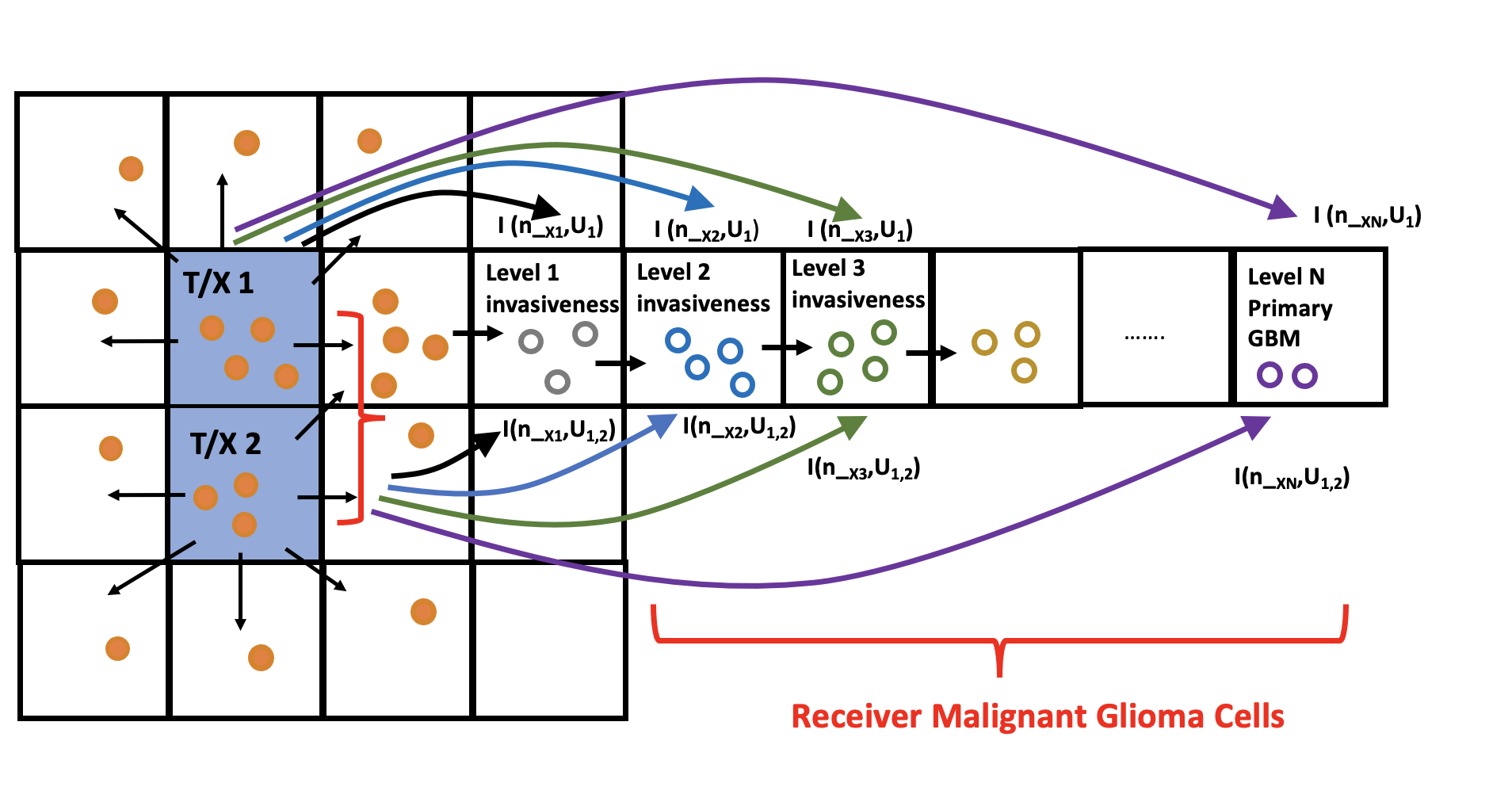}
 \caption{Mutual information Comparison Scenarios- Go}
\label{s3sq}
 \end{center}
 \end{figure}
 
   \begin{figure}[H]
 \begin{center}
\includegraphics[trim=0cm 0cm 0cm 0cm ,clip=true, width=0.6\columnwidth]{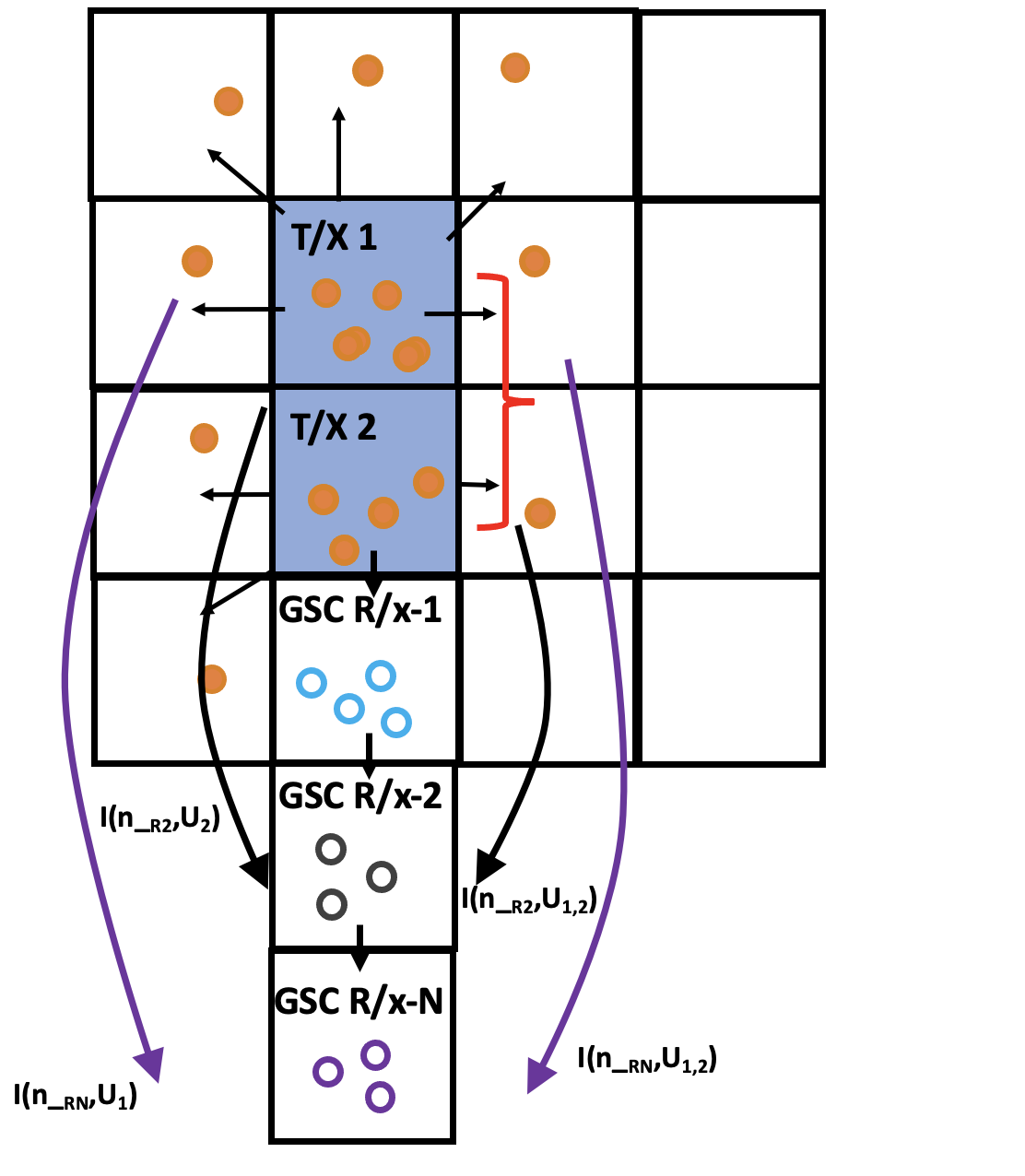}
 \caption{Mutual information Comparison Scenarios- Grow}
\label{s3s}
 \end{center}
 \end{figure}
In Figures 6 and 7 we present various comparison scenarios for mutual information of single and multiple transmitter glioma stem cells in both Grow and Go configurations for an increasing number of receiver malignant glioma cells. As shown in Figures 8 and 9 the total mutual information (for both Go and Grow) tends to increase as the number of both transmitters and receiver malignant glioma cells increase. This is due to the reduction in noise as a result of increased capture of molecules by receivers. This result also suggests that a higher mutual information among glioma cells in a particular direction will result in higher GBM growth rate (stem growth) in that direction as shown in Figure 1.

%   \begin{figure}  [H]
% \begin{center}
% \includegraphics[trim=0cm 0cm 0cm 0cm ,clip=true, width=0.8\columnwidth]{1xn.eps}
% \caption{ Impact of Multiple Cells on Mutual Information}
% \label{molecular communications-n}
% \end{center}
% \end{figure}

% \begin{figure} [H]
% \begin{center}
% \includegraphics[trim=0cm 0cm 0cm 0cm ,clip=true, width=0.8\columnwidth]{1xm.eps}
% \caption{Impact of Multiple Cells on Propagation Speed}
% \label{molecular communications-n2}
% \end{center}
% \end{figure}

% First in Figure 4 we show the mutual information of the proposed model for multiple transmitter and increasing number of receiver malignant glioma  cells. As depicted the mutual information is increasing steadily for various configurations of cells and  increasing number of receiver cells. Intuitively this trend can be explained by the increase in the information transfer between cells resulting in higher mutual information. Furthermore we learn that for the hybrid configuration of cells the loss of molecules in the propagation medium is slightly lower resulting in higher mutual information as compared to other two configurations. 

\begin{figure}[ht]
\begin{center}
\includegraphics[trim=0cm 0cm 0cm 0cm ,clip=true, width=0.8\columnwidth]{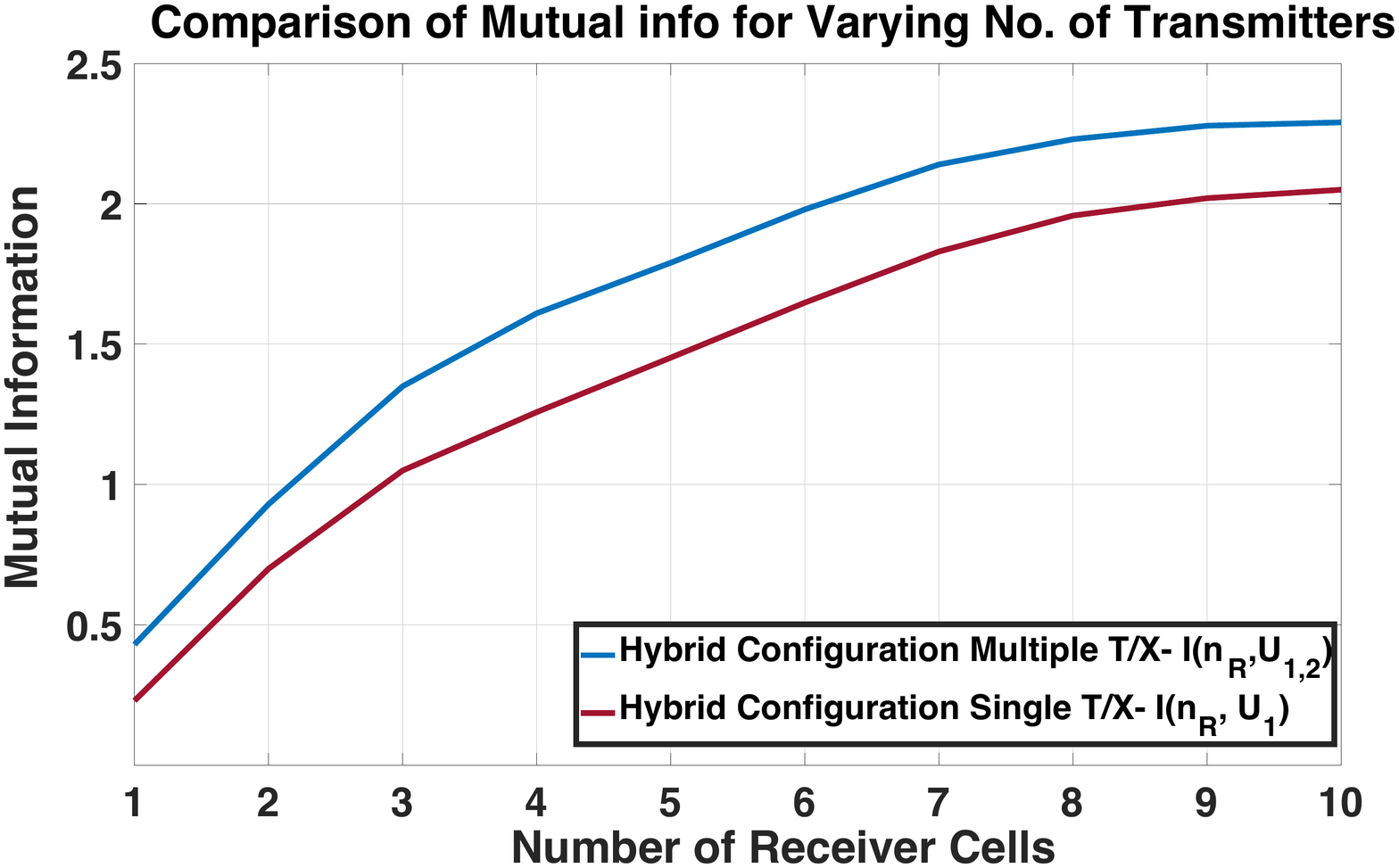}
\caption{Mutual information-Single vs Multiple Transmitter- Go}
\label{molecular communications-n31}
\end{center}
\end{figure}

\begin{figure}[ht]
\begin{center}
\includegraphics[trim=0cm 0cm 0cm 0cm ,clip=true, width=0.8\columnwidth]{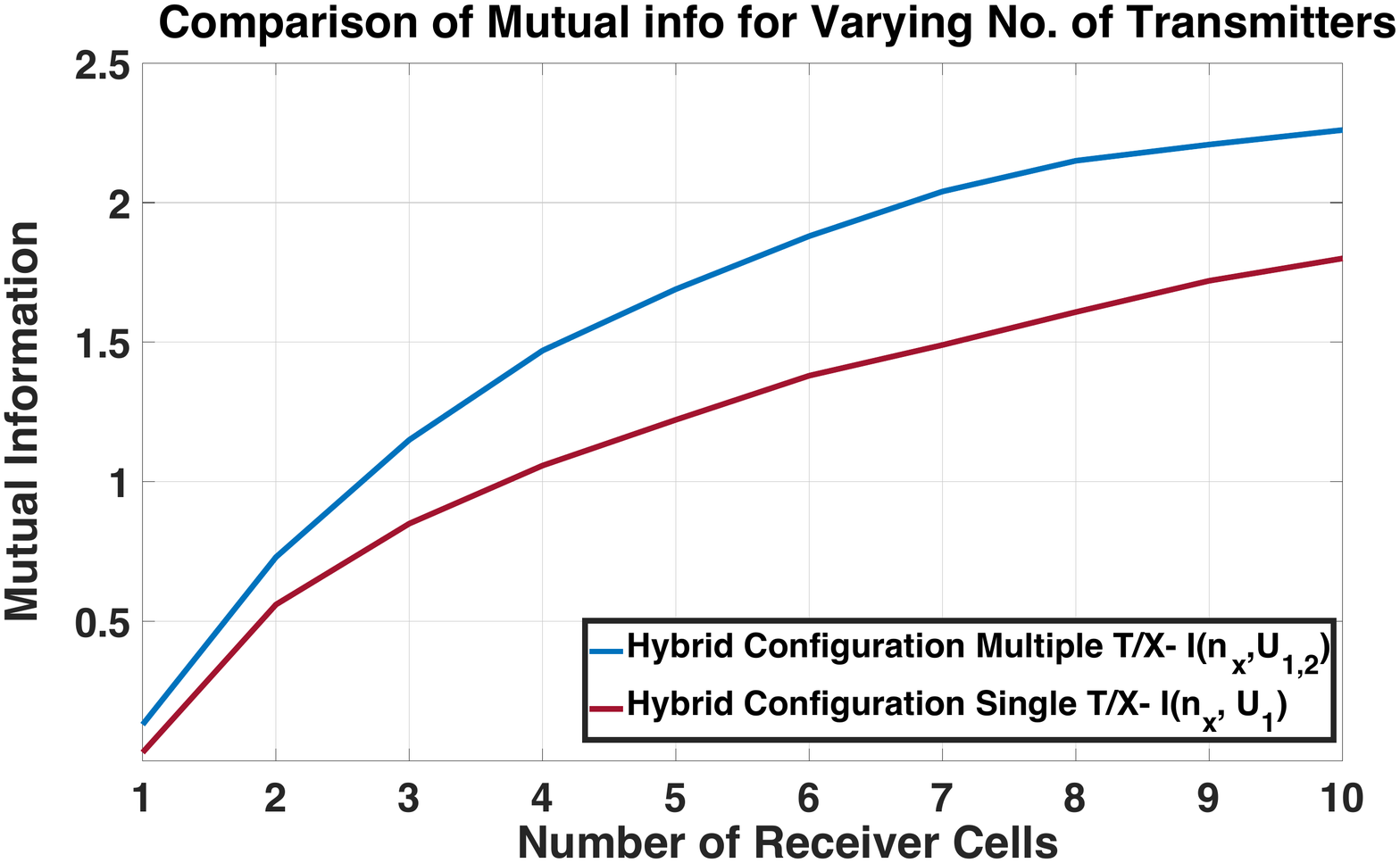}
\caption{Mutual information-Single vs Multiple Transmitter- Grow}
\label{molecular communications-n32}
\end{center}
\end{figure}

Finally in Figure 10 we present a comparison for information propagation between node and stem cells of GBM (as shown in Figure 1). We learn that due to lower density of cells in the stem the information propagation speed is higher in stem as compared to node. However, we note that that due to lower density of cells in stem this area can be targeted by potential therapeutic mechanisms for GBM.

 \begin{figure}[ht]
 \begin{center}
\includegraphics[trim=0cm 0cm 0cm 0cm ,clip=true, width=0.75\columnwidth]{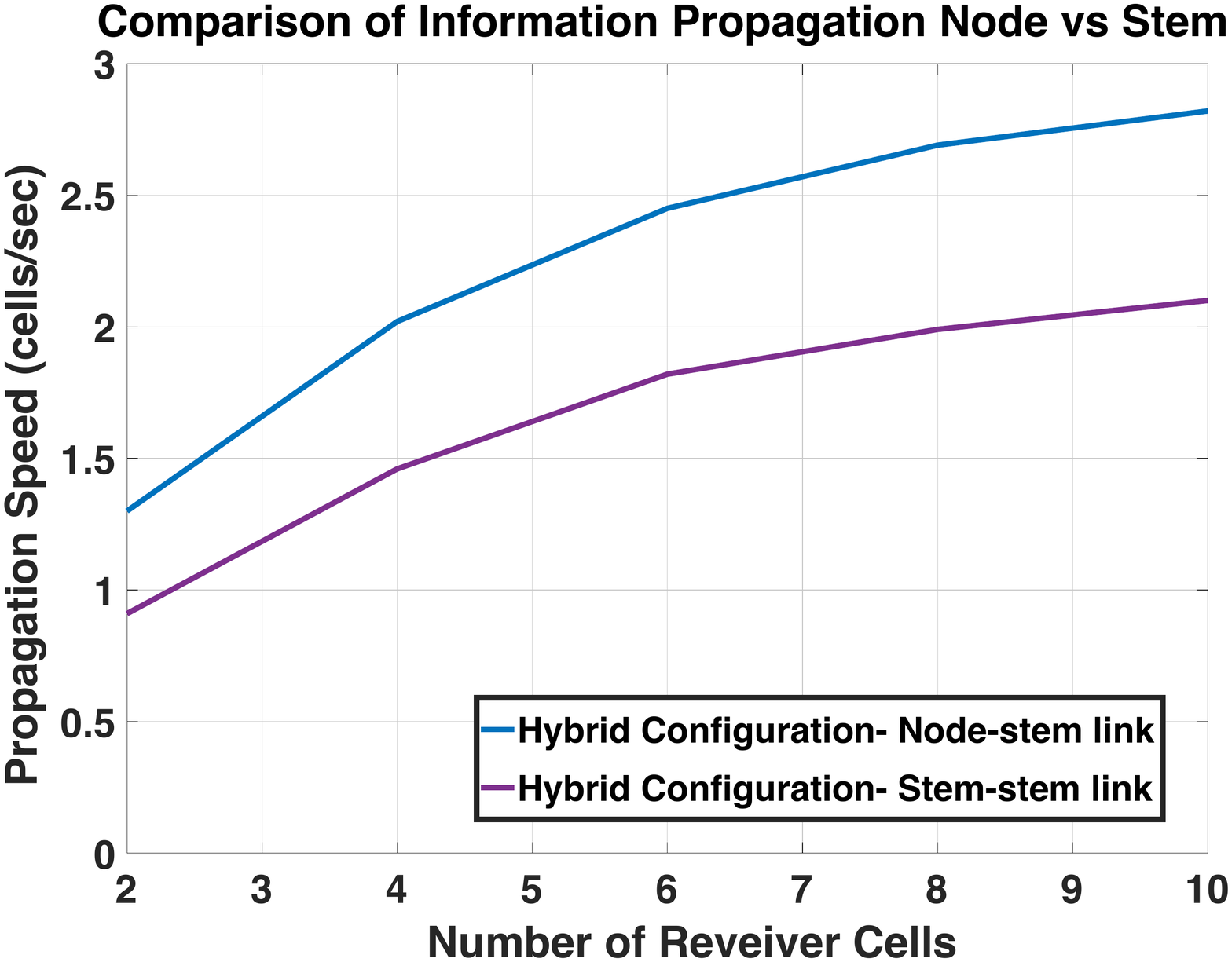}
 \caption{Information propagation speed comparison- Node vs Stem}
\label{s3}
 \end{center}
 \end{figure}

 \section{Conclusion}

 This paper is focused on understanding the mechanisms by which a network of multiple bio-nanomachines (transmitters and receivers) in a molecular communication system influence the growth structure of GBM tumour, a grade IV malignant glioma. By using a voxel model for multiple transmitter and receiver cells, this paper provides new insights into the role of inter-cellular communication (i.e. the mutual information between cells) in the evolution and progression of GBM (stem from node).  From the results of this paper we learn that the growth of tumour in any particular direction is driven by the increase in the mutual information between multiple cells in the node-stem structure. We further learn that information propagation speed between cells can vary at different points in the node and stem. This knowledge can be useful for developing future therapeutic mechanisms targeting GBM.

\ifCLASSOPTIONcaptionsoff
  \newpage
\fi

\bibliographystyle{ieeetr}
\bibliography{nano2017,book,nano2018}

% that's all folks
\end{document}